\documentclass[prc,twocolumn, amsmath,amssymb,floatfix,nofootinbib]{revtex4}
\usepackage{mathrsfs,bm}
\usepackage{longtable,lscape}
\usepackage{txfonts}
\usepackage{indentfirst}
\usepackage{graphicx,color,dcolumn,booktabs}
\usepackage{multirow}
\usepackage{indentfirst}
\usepackage{multirow}
\usepackage{epsfig}
\usepackage{graphicx,color,dcolumn}
\usepackage{epstopdf}
\usepackage{amsmath}
\usepackage{diagbox}
\usepackage{changes}
\usepackage{threeparttable}
\usepackage{booktabs}
\usepackage{color}
\usepackage{amssymb}
\usepackage{cancel}
\definecolor{cover}{rgb}{0.77,0.87,0.88}
\definecolor{blueone}{rgb}{0.1,0.1,.7}
\definecolor{citec}{rgb}{0.14,0.47,0.09}
\definecolor{two}{rgb}{0.0,0.5,0.}
\definecolor{three}{rgb}{.5,.1,0.15}
\usepackage[bookmarks=true,bookmarksopen=false,plainpages=false,breaklinks=true,
   bookmarksnumbered=true,hypertexnames=false,
   filecolor=blue,urlcolor=three,menucolor=three,
   linkcolor=three,citecolor=blueone, colorlinks,
   anchorcolor=blue,runcolor=pink,frenchlinks=red
   pdfstartview=FitH,pdftitle=title,%
   pdfauthor=author]{hyperref}

\usepackage{relsize}
\usepackage{xspace}
\def\babar{\mbox{\slshape B\kern-0.1em{\smaller A}\kern-0.1em
    B\kern-0.1em{\smaller A\kern-0.2em R}}}

\newcolumntype{C}{>{$}c<{$}}
\AtBeginDocument{
\heavyrulewidth=.08em
\lightrulewidth=.05em
\cmidrulewidth=.03em
\belowrulesep=.65ex
\belowbottomsep=0pt
\aboverulesep=.4ex
\abovetopsep=0pt
\cmidrulesep=\doublerulesep
\cmidrulekern=.5em
\defaultaddspace=.5em
}

\allowdisplaybreaks[4]

\begin{document}
\title{Role of $\Sigma^*(1/2^-)$ baryons in $\Lambda_c\to \Lambda\pi^+\pi^+\pi^-$  decay}
\author{Jun He}
\email{junhe@njnu.edu.cn}
\affiliation{School of Physics and Technology, Nanjing Normal University, Nanjing 210097, China}

\date{\today}
\begin{abstract}

Based on Belle data, the structure around 1435 MeV and the role of the $\Sigma(1380)$ resonance in the decay process $\Lambda_c \to \Lambda \pi^- \pi^+ \pi^+$ are investigated.  The $\Lambda \pi^\pm$ invariant mass spectra are
approximately reproduced by the contribution from the $\Sigma(1385)$, modeled as
a Breit-Wigner resonance combined with background. However, an additional
contribution is required to account for the pronounced enhancement near 1435 MeV.
To describe this structure, two scenarios are considered for the $\Sigma(1435)$:
one as a Breit-Wigner resonance with spin-parity $1/2^-$ described via effective
Lagrangians, and the other as a rescattering effect arising from coupled-channel
interactions. Specifically,  the channels $\bar{K}^0p$,
$\pi^0\Sigma^+$, $\pi^+\Sigma^0$, $\pi^+\Lambda$, and $\eta\Sigma^+$ are included for the
positively charged mode, and $K^-n$, $\pi^0\Sigma^-$, $\pi^-\Sigma^0$,
$\pi^-\Lambda$, and $\eta\Sigma^-$ for the negatively charged mode. The
rescattering contributions are evaluated using the quasipotential Bethe-Salpeter
equation approach.  Simulated invariant mass spectra are generated and fitted to
the experimental data. Both the Breit-Wigner and rescattering mechanisms provide
reasonable descriptions of the enhancement near 1435 MeV. The rescattering approach
naturally generates a cusp-like structure near threshold, which closely
resembles the feature observed in the data. Furthermore, the inclusion of the
$\Sigma(1380)$ resonance with spin-parity $1/2^-$ significantly improves the fit
in the low-energy region and further enhances the description near 1435 MeV
through interference effects. These results underscore the essential roles of
the two $\Sigma^*(1/2^-)$ states in accurately describing the invariant mass
distributions in the decay $\Lambda_c \to \Lambda \pi^+ \pi^+ \pi^-$.

\end{abstract}

\maketitle
\section{Introduction}

Light baryon resonances provide a crucial window into the nonperturbative
regime of quantum chromodynamics, as their masses, decay patterns, and
production mechanisms encode the underlying dynamics of strong interactions.
Thanks to advances in accelerator and detector technologies, a wealth of
high-precision data has been accumulated from facilities such as JLab and
J-PARC, leading to an increasingly complete mapping of the nucleon and $\Delta$
resonance spectra. In contrast, hyperon spectroscopy remains far less explored:
although $\Sigma$ resonances were first reported in the 1960s, the Review of Particle Physics (PDG) currently lists only nine firmly established states (three- or
four-star ratings), despite lattice QCD calculations and quark model predictions
suggesting the existence of approximately 70 such states~\cite{ParticleDataGroup:2024cfk,Capstick:1986ter}. Consequently,
significant experimental and theoretical efforts have been devoted in recent
years to advancing our understanding of hyperon spectrum.

In recent years, $\Sigma$ resonances with spin-parity $1/2^-$ have attracted
considerable attention. Within the framework of the constituent quark model, the
lowest $\Sigma(1/2^-)$ state is predicted to have a mass around 1630
MeV~\cite{Capstick:1986ter}. Analyses of $K^-$-induced reaction data suggest the
observation of a resonance $\Sigma(1620)$, potentially corresponding to this
predicted state~\cite{Sarantsev:2019xxm}. However, this resonance is currently
rated as a one-star state by PDG, reflecting a low
level of experimental confirmation~\cite{ParticleDataGroup:2024cfk}.
Interestingly, several theoretical studies have predicted the existence of
lower-mass $\Sigma(1/2^-)$ states, and experimental signals consistent with such
states have also been reported, further stimulating interest and ongoing
investigation in this sector~\cite{Oller:2000fj,Jido:2003cb}.

A $\Sigma(1/2^-)$ state around 1380 MeV has been theoretically predicted and
supported by several experimental analyses. In the diquark model proposed by
Jaffe and Wilczek, a $\Sigma(1/2^-)$ resonance near 1360 MeV was
suggested~\cite{Zhang:2004xt}. Wu et al. found evidence for a new
$\Sigma(1/2^-)$ resonance in the $K^-p \to \Lambda\pi^+\pi^-$ reaction,
characterized by a broader width than the well-known $\Sigma(1385)$ with
spin-parity $3/2^-$\cite{Wu:2009tu,Wu:2009nw,Huwe:1969te}.
Reference~\cite{Xie:2014zga} demonstrated that the near-threshold enhancement in the
$\pi^0\Lambda$ invariant mass spectrum of $\Lambda p \to \Lambda p \pi^0$ can be
well described by including this $\Sigma(1/2^-)$ state. Further support for its
existence comes from additional theoretical and experimental studies, including
photoproduction $\gamma N \to K^+\pi\Lambda$, $K^-p$ scattering, and
neutrino-induced reactions~\cite{Chen:2013vxa,Roca:2013av,Gao:2010hy}. In
$\Lambda_c^+ \to \eta\pi^+\Lambda$ decays, discrepancies observed in the Dalitz
plot may also originate from this state~\cite{Wang:2022nac}, and its inclusion
significantly improves the fit to Belle data~\cite{Lyu:2024qgc}. Similarly, in
$\Lambda_c^+ \to p \bar{K}^0 \eta$, the near-threshold enhancement in the
$p\bar{K}^0$ mass spectrum is well reproduced by incorporating this low-lying
$\Sigma(1/2^-)$ state, leading to better agreement with Belle
measurements~\cite{Li:2024rqb,Pavao:2018wdf}. Altogether, these findings provide
strong evidence for the existence of a $\Sigma(1/2^-)$ resonance near 1380 MeV.

In addition to the $\Sigma(1/2^-)$ resonance around 1380 MeV, another
$\Sigma(1/2^-)$ state located closer to the $\bar{K}N$ threshold has been
proposed in various theoretical frameworks. The $\Lambda(1405)$ is widely
interpreted as a $\bar{K}N$ bound state with dominant isoscalar components~\cite{Oller:2000fj,Jido:2003cb,Oset:1997it,Mai:2014xna,Mai:2020ltx,Sadasivan:2022srs}, and
the potential existence of its isovector partner has also attracted considerable
attention. For instance, the chiral unitary approach in Ref.~\cite{Jido:2003cb}
revealed two poles at $1401-40i$ MeV and $1488-114i$ MeV. More recent analyses
based on covariant baryon chiral perturbation theory up to
next-to-next-to-leading order~\cite{Wang:2024jyk} reinforced this two-pole
picture, identifying a narrow pole at $1432 - 18i$ MeV, manifesting as a
cusp-like structure near the $\bar{K}N$ threshold, along with a broader pole at
$1364 - 110i$ MeV. Similarly, the coupled-channel calculations in
Ref.~\cite{Khemchandani:2018amu} predicted both a narrow, higher-mass
$\Sigma(1/2^-)$ state and a broader, lower-mass partner. These theoretical
predictions receive additional support from the CLAS data on the $\gamma p \to K
\Sigma \pi$ reaction~\cite{Roca:2013cca, CLAS:2013rjt}. Moreover,
Ref.~\cite{Xie:2018gbi} studied the decay $\Lambda_c^+ \to \pi^+ \pi^0 \pi^-
\Sigma^+$ via a triangle loop mechanism, suggesting that a narrow
$\Sigma(1/2^-)$ resonance with mass around 1430 MeV could manifest as a distinct
peak near the $\bar{K}N$ threshold.

Recently, the Belle Collaboration reported a prominent structure around
1435~MeV, located near the $\bar{K}N$ threshold, in the $\Lambda\pi^+$ and
$\Lambda\pi^-$ invariant mass spectra from the decay $\Lambda_c^+ \to
\Lambda\pi^+\pi^+\pi^-$~\cite{Belle:2022ywa}. In the present study, this structure is referred to as  $\Sigma(1435)$. According to Belle's analysis, the data can be
described using Breit-Wigner parametrizations, yielding a mass and width of
$1434.3 \pm 0.6\ (\text{stat}) \pm 0.9\ (\text{syst})$ MeV and $11.5 \pm 2.8\
(\text{stat}) \pm 5.3\ (\text{syst})$ MeV in the $\Lambda\pi^+$ channel, and
$1438.5 \pm 0.9\ (\text{stat}) \pm 2.5\ (\text{syst})$ MeV with a broader width
of $33.0 \pm 7.5\ (\text{stat}) \pm 23.6\ (\text{syst})$ MeV in the
$\Lambda\pi^-$ channel. Since both structures lie very close to the $\bar{K}N$
threshold, the possibility that they originate from a cusp effect was also
considered. However, due to the limited statistical precision, the experimental
analysis could not definitively discriminate between a genuine resonance and a
threshold cusp. Nevertheless, the observed structure appears to be related to
the longstanding features near the $\bar{K}N$ threshold that have been noted and
discussed in the literature.

In the experimental analyses, the spin-parity of the observed structures was not
conisdered. Nevertheless, the positions of these structures and their decays
into $\Lambda\pi^\pm$ suggest they are consistent with a $\Sigma(1/2^-)$
assignment, as widely discussed in the literature. Notably, the experimental
focus was mainly on the region near the $\bar{K}N$ threshold, and data below
1380 MeV were not included in the fit. Even the parameters of the
well-established $\Sigma(1385)$ resonance were not treated with full rigor. In
fact, a possible $\Sigma(1/2^-)$ resonance near 1380 MeV could interfere with
the $\Sigma(1385)$, significantly complicating the interpretation and
description of the complete $\Lambda\pi$ invariant mass spectrum.

Motivated by these considerations, the present study investigates the roles of
both the $\Sigma(1380)$ and $\Sigma(1435)$ states with spin-parity $1/2^-$ in
describing the Belle data on the $\Lambda\pi^\pm$ invariant mass spectra. As in
the experimental analyses, the structure $\Sigma(1435)$ near the $\bar{K}N$
threshold is modeled using both a Breit-Wigner resonance and a rescattering
cusp. However, in contrast to the experimental treatment, the spin-parity of
$1/2^-$ is explicitly imposed for the $\Sigma(1435)$.  The Breit-Wigner
resonance contributions are described using effective Lagrangian methods, while
the rescattering effects are incorporated through coupled-channel interactions:
$\bar{K}^0 p$, $\pi^0 \Sigma^+$, $\pi^+ \Sigma^0$, $\pi^+ \Lambda$, and $\eta
\Sigma^+$ for the positively charged mode, and $K^- n$, $\pi^0 \Sigma^-$, $\pi^-
\Sigma^0$, $\pi^- \Lambda$, and $\eta \Sigma^-$ for the negatively charged mode.
These interactions are treated within the framework of the quasipotential
Bethe-Salpeter equation (qBSE) approach.  Furthermore, unlike the experimental
analyses that fitted the $\Lambda\pi^+$ and $\Lambda\pi^-$ spectra separately
and neglected interference effects, the present work performs a simultaneous fit
to both spectra and explicitly includes the interference among different
contributions.

The remainder of this paper is organized as follows. Section~\ref{Sec:
Formalism} presents the theoretical framework, including the decay mechanisms
considered for $\Lambda_c \to \Lambda \pi^+ \pi^+ \pi^-$, the event simulation
procedure, the effective Lagrangians used to describe Breit-Wigner resonances,
the formulation of the interaction potential kernel, and a brief overview of the
qBSE approach employed to account for
rescattering effects. Section~\ref{Sec: results} provides the numerical results
obtained from various fitting strategies. The roles of the $\Sigma(1435)$,
treated either as a Breit-Wigner resonance or as a dynamically generated state
via rescattering, along with the contribution of the additional $\Sigma(1380)$
resonance, are analyzed in detail. Finally, the main findings and their
implications are summarized in Section~\ref{Summary}.

\section{Theoretical frame}\label{Sec: Formalism}

In this section, the mechanism for the decay process $\Lambda_c^+ \to
\Lambda\pi^+\pi^+\pi^-$ is first introduced. In the present analysis, the
$\Sigma(1385)$ and $\Sigma(1380)$ are modeled as Breit-Wigner
resonances with spin-parities $3/2^+$ and $1/2^-$, respectively, using effective
Lagrangian formulations. In addition to the Breit-Wigner treatment, the
$\Sigma(1435)$ structure is considered as a dynamically generated state arising
from coupled-channel interactions, studied within the framework of the qBSE
approach.

\subsection{Decay mechanism of $\Lambda_c^+ \to \Lambda\pi^+\pi^+\pi^-$}

\begin{figure}[h!]
  \includegraphics[bb=320 50 1400 650,clip,scale=0.33]{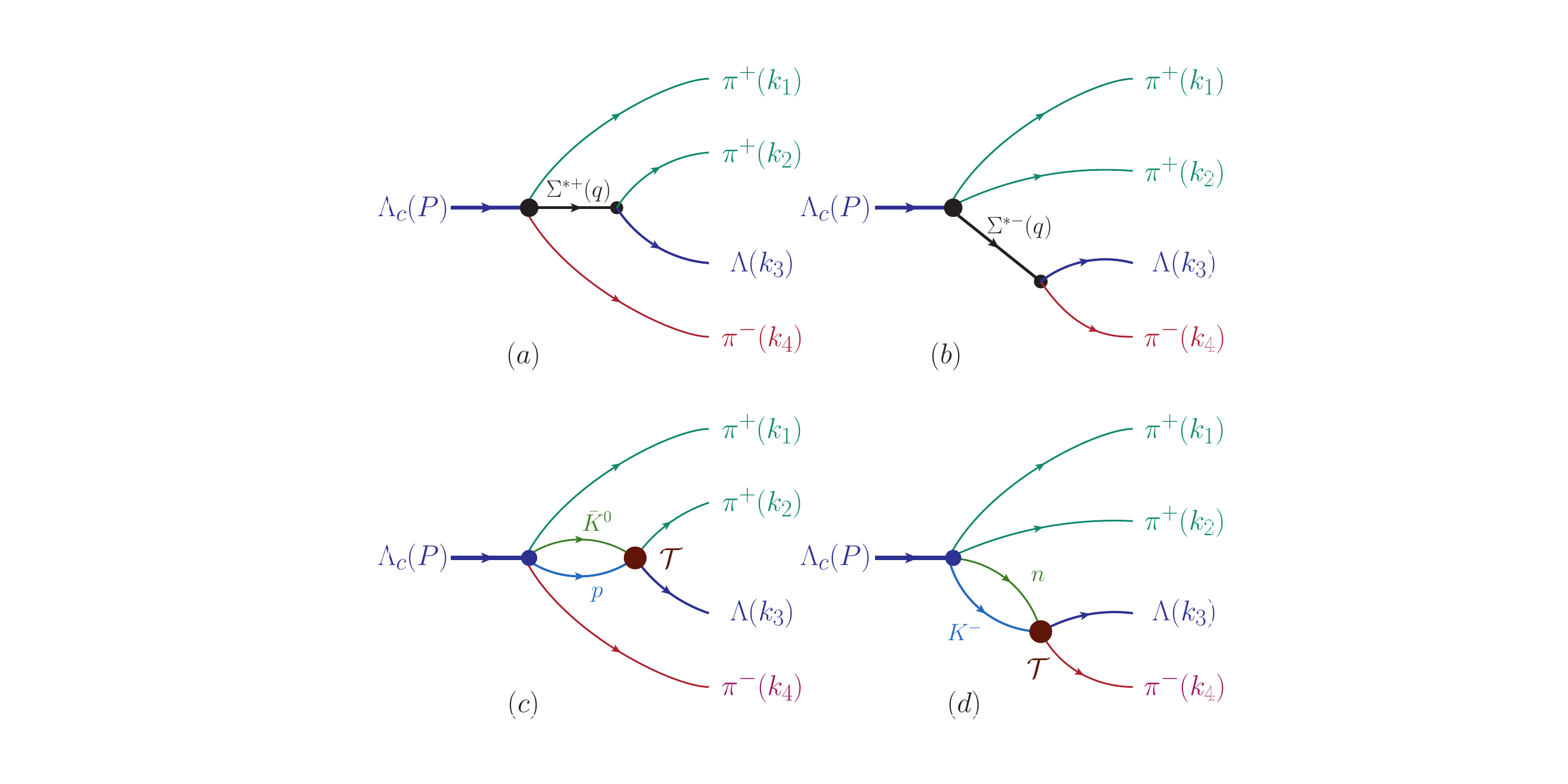}
  \caption{Feynman diagrams for the decay $\Lambda_c^+ \to \Lambda \pi^+ \pi^+ \pi^-$. Panels (a) and (b) illustrate mechanisms involving intermediate Breit-Wigner resonances contributing to the $\Lambda \pi^+$ and $\Lambda \pi^-$ final states, respectively. Panels (c) and (d) show the corresponding mechanisms involving coupled-channel interactions for the positive and negative charge $\Lambda\pi^\pm$ systems. The brown filled circles denote the coupled-channel amplitudes $\mathcal{T}$.}
  \label{diagram}
\end{figure}

The Belle data exhibit a prominent peak structure around 1380~MeV and a smaller enhancement near 1435~MeV. To describe these structures, two types of decay mechanisms are considered in the present work, as illustrated in Fig.~\ref{diagram}.

In Fig.~\ref{diagram} (a) and (b), the decay mechanisms involving intermediate
Breit-Wigner resonances are illustrated. In this mechanism, the $\Lambda_c^+$
first decays into a $\Sigma^*$ resonance accompanied by two pions, and then the
intermediate $\Sigma^*$ subsequently decays into a $\Lambda$ baryon and a pion.
The intermediate resonance can carry either positive or negative charge,
decaying into $\Lambda\pi^+$ and $\Lambda\pi^-$ final states, as shown in
Fig.~\ref{diagram} (a) and (b), respectively.

In the present work, the Breit-Wigner formalism is used to describe the
$\Sigma(1380)$, $\Sigma(1385)$, and $\Sigma(1435)$. However, an alternative
approach is also adopted, where the $\Sigma(1435)$ is treated as a dynamically
generated structure arising from coupled-channel interactions, as illustrated in
Fig.~\ref{diagram} (c) and (d). In this mechanism, the $\Lambda_c^+$ first
decays into two pions and a $\bar{K}N$ pair. Since the primary focus is on the $\Sigma(1435)$ structure, only the $\bar{K}N$ channel is considered in the initial $\Lambda_c^+$ decay. Similar to the Breit-Wigner case, rescattering
effects can occur in both the positive and negative charge channels. In the
positive charge channel, the $\bar{K}^0p$ system connects to the final
$\Lambda\pi^+$ state through a coupled-channel rescattering process, involving
channels $\bar{K}^0p$, $\pi^0\Sigma^+$, $\pi^+\Sigma^0$,
$\pi^+\Lambda$, and $\eta\Sigma^+$. The negative charge channel proceeds
analogously.

The total amplitude for the decay can be expressed as
\begin{align}
{\cal M} = \sum_{\Sigma^{*\pm}} \left( e^{i2\pi\phi_{\Sigma^{*+}}} {\cal M}_{\Sigma^{*+}} + e^{i2\pi\phi_{\Sigma^{*-}}} {\cal M}_{\Sigma^{*-}} \right) + {\cal M}_{\text{bk}},
\end{align}
where $e^{i2\pi\phi_{\Sigma^{*\pm}}}$ and ${\cal M}_{\Sigma^{*\pm}}$ denote the relative phases and amplitudes associated with the charge positive and negative resonance $\Sigma^{*\pm}$, respectively. This notation can also be used when describing contributions from rescattering mechanisms.

In the present analysis,  following the experimental procedure, the two identical $\pi^+$ mesons are treated as indistinguishable. Consequently, the amplitudes involving the positive charge channel must be symmetrized as
\begin{align}
{\cal M}_{\Sigma^{*+}} = \frac{1}{\sqrt{2}} \left[ {\cal M}_{\Sigma^{*+}}(1234) + (2 \leftrightarrow 1) \right],
\end{align}
where ${\cal M}_{\Sigma^{*+}}(1234)$ represents the amplitude corresponding to the diagrams shown in Fig.~\ref{diagram}(a) or \ref{diagram}(c), with the indices (1234) labeling the final particles. For the negative charge channel, the two $\pi^+$ mesons have already been symmetrized in the definition of the amplitude, and thus no additional treatment is required.

All other contributions are treated as a constant background term, given by
\begin{align}
i{\cal M}_{\text{bk}} = g_{\text{bk}}\bar{u}_\Lambda u_{\Lambda_c},
\end{align}
where $g_{\text{bk}}$ is a free parameter and $u_{\Lambda_c}$ and $u_\Lambda$ are the Dirac spinors for the $\Lambda_c^+$ and $\Lambda$ baryons, respectively.
Additionally, to account for possible trends in the higher invariant mass region of the $\Lambda\pi^\pm$ spectra,  a first-order Chebyshev polynomial modification of the background is also considered in the current work, written as $1 + cM$, where $M$ is the invariant mass of the $\Lambda\pi^\pm$ system. Considering that the phase space factor is inlcuded explicitly in the current treatment, a first-order polynomial is adopted instead of a second-order one, which is different from that adopted in the Belle analyses~\cite{Belle:2022ywa}.

With the decay amplitude, the differential decay width can be written as \begin{align} d\Gamma = \frac{1}{2E} \sum |{\cal M}|^2 d\Phi, \end{align} where $E$ is the energy of the initial $\Lambda_c^+$ baryon, and $d\Phi$ represent the Lorentz-invariant phase space elements.
In this work, the phase space element $d\Phi$ is simulated using the method proposed in Ref.~\cite{James:1968gu}, implemented in the Julia programming language. The corresponding Julia package is available on GitHub~\cite{code}. A total of $10^8$ events were generated for the simulation. The momenta of the final-state particles are sampled using a Monte Carlo approach that strictly enforces energy–momentum conservation. From these simulated momenta, the invariant mass spectra of the $\Lambda\pi^\pm$ systems are constructed and subsequently used in the analysis.

\subsection{Amplitudes with intermediate Breit-Wigner resonances}

The amplitudes for the diagrams involving Breit-Wigner resonances [see Fig.~\ref{diagram} (a) and (b)] are derived using standard Feynman rules based on effective Lagrangians. In this work, the $\Sigma(1385)$ and $\Sigma(1380)$ are all treated as Breit-Wigner resonances, and this is also one of the two scenarios considered for the $\Sigma(1435)$. Among these, the $\Sigma(1385)$ is assigned a spin-parity of $3/2^+$, while both the $\Sigma(1380)$ and $\Sigma(1435)$ are considered to have spin-parity $1/2^-$.

For the intermediate $\Sigma^*$ resonance with $J^P = 3/2^+$, the effective Lagrangians describing the $\Lambda_c$ decay into $\Sigma(3/2^+)$ plus two pions, and the subsequent $\Sigma(3/2^+)$ decay into $\Lambda\pi$  are written as~\cite{Ahn:2019rdr,Wang:2022xqc}
\begin{align}
{\cal L}_{\Lambda_c\Sigma(3/2^+)\pi\pi}&=\frac{i}{m_\pi}\bar{\Sigma}_\mu(\pi\partial^\mu \pi'+\pi'\partial^\mu \pi)(g^{PV}_{\Sigma(3/2^+)}-g_{\Sigma(3/2^+)}^{PC}\gamma_5)\Lambda_c,\nonumber\\
{\cal L}_{\Sigma(3/2^+)\Lambda\pi}&=-\frac{ig_{\Sigma(3/2^+)}}{m_\pi}\bar{\Lambda}\partial^\mu\pi\Sigma_\mu,
\end{align}
where $\pi$ and $\pi'$ denote the two pions, and $g^{PV}_{\Sigma(3/2^+)}$ and $g^{PC}_{\Sigma(3/2^+)}$ are the coupling constants for parity-violating and parity-conserving interactions, respectively. Since it is a weak decay, these couplings may differ for $\Sigma(3/2^+)$ states of different charges.

The propagator for a spin-$3/2$ partilce is taken as  \begin{align}
G_{\mu\nu}(q)&=\frac{i}{q^2-m^2+i m\Gamma}(q\mkern -9.5 mu /+m)[-g_{\mu\nu}\nonumber\\
&+\frac{1}{3}\gamma_\mu\gamma_\nu+\frac{1}{3m}(\gamma_\mu q_\nu-\gamma_\nu q_\mu)+\frac{2}{3m^2}q_\mu q_\nu],
\end{align}
where $q$ and $m$ are the four-momentum and mass of the spin-$3/2$ partilce, and $\Gamma$ is its decay width.

Following the Feynman rules, the corresponding amplitude can be obtained as 
\begin{align}
i{\cal M}_{\Sigma(3/2^+)^{\pm}}=g_{\Sigma^*(3/2^+)} \frac{\bar{\Lambda}\Delta^\pm_3(1-\gamma_5)\Lambda_c}{q^2-m^2_{\Sigma(3/2^+)^{\pm}}+im_{\Sigma(3/2^+)^{\pm}}\Gamma_{\Sigma(3/2^+)^{\pm}}},
\end{align}
with
\begin{align}
\Delta^\pm_3&=(q\mkern -9.5 mu /+m_{\Sigma(3/2^+)^{\pm}})[-k\cdot p+\frac{1}{3}k\mkern -9.5 mu /{p\mkern -9.5 mu /}\nonumber\\
&+\frac{1}{3m_{\Sigma(3/2^+)^{\pm}}}({k\mkern -9.5 mu /} q\cdot p-{p\mkern -9.5 mu /} q\cdot k)+\frac{2}{3m^2_{\Sigma(3/2^+)^{\pm}}}q\cdot k q\cdot p],
\end{align}
where $p = k_1 + k_4$ or $k_1 + k_2$ corresponds to the cases of charge positive and negative $\Sigma(3/2^+)$, respectively.
Following the literature, the coupling constants are set equal, i.e., $g^{PV}_{\Sigma(3/2^+)} = g^{PC}_{\Sigma(3/2^+)}$, for simplicity~\cite{Ahn:2019rdr,Wang:2022xqc,Kong:2024scz}. This treatment simplifies the fitting process by reducing the number of free parameters while effectively capturing the strength of the $\Sigma(3/2^+)$ contributions. In the fitting procedure, the coupling constants are absorbed into an effective parameter, defined as \begin{align} g_{\Sigma(3/2^+)^\pm} = -\frac{g_{\Sigma(3/2^+)} g^{PC}_{\Sigma(3/2^+)^\pm}}{m_\pi^2}. \end{align}

If the intermediate $\Sigma^*$ states have spin-parity $1/2^-$, the effective Lagrangians describing the $\Lambda_c$ decay and the subsequent $\Sigma(1/2^-)$ decay can be written as~\cite{Ahn:2019rdr}
\begin{align} {\cal L}_{\Lambda_c\Sigma(1/2^-)\pi\pi} &= i\bar{\Sigma} \pi \pi (g^{PV}_{\Sigma(1/2^-)} - g^{PC}_{\Sigma(1/2^-)}  \gamma_5) \Lambda_c, \nonumber\\
{\cal L}_{\Sigma(1/2^-)\Lambda\pi} &= -ig_{\Sigma(1/2^-)}  \bar{\Lambda} \pi \Sigma, \end{align} 
where $g^{PV}_{\Sigma(1/2^-)} $ and $g^{PC}_{\Sigma(1/2^-)} $ are the coupling constants associated with parity-violating and parity-conserving interactions, respectively, and are also taken to be equal.

Following the standard Feynman rules, the decay amplitude can be obtained as 
\begin{align} i{\cal M}_{\Sigma(1/2^-)^{\pm}} = g_{\Sigma(1/2^-)^\pm} \frac{\bar{\Lambda} \Delta^\pm (1-\gamma_5) \Lambda_c}{q^2 - m^2_{\Sigma(1/2^-)^{\pm}} + i m_{\Sigma(1/2^-)^{\pm}} \Gamma_{\Sigma(1/2^-)^{\pm}}}, 
\end{align} 
where $\Delta$ is given by $\Delta^\pm=(q\mkern -9.5 mu /+m_{\Sigma(1/2^-)^{\pm}})$. In the fitting procedure, the combination of coupling constants is also treated as a single effective parameter, defined as $g_{\Sigma(1/2^-)^\pm} = g_{\Sigma(1/2^-)} g^{PC}_{\Sigma(1/2^-)^\pm} $.

\subsection{Amplitudes with rescattering}

For the structure observed around 1435~MeV,  two scenarios, a
Breit-Wigner resonance and a coupled-channel rescattering process, are considered in the current work. In the
rescattering mechanism, the coupled-channel interactions among the channels $\bar{K}^0p$,
$\pi^0\Sigma^+$, $\pi^+\Sigma^0$, $\pi^+\Lambda$, and $\eta\Sigma^+$ for the
positively charged mode, and $K^-n$, $\pi^0\Sigma^-$, $\pi^-\Sigma^0$,
$\pi^-\Lambda$, and $\eta\Sigma^-$ for the negatively charged mode, are taken into account.
These interactions correspond to the isovector counterpart of the dynamics
associated with the $\Lambda(1405)$ resonance, and have been established in
Ref.~\cite{Oset:1997it}.  The interaction kernel between different channels is given by 
\begin{align} i{\cal V}^{kl}_{\lambda'\lambda} = -C^{kl} \frac{1}{4f^2} \bar{u}_{\lambda'}(p)\gamma^\mu u_\lambda(p') (k_\mu + k'_\mu), \end{align} 
where $u_{\lambda'}(p')$ and $\bar{u}_\lambda(p)$ are the Dirac spinors for the incoming and outgoing baryons, respectively, while $k^{(')}$ and $p^{(')}$ denote the momenta of the mesons and baryons involved in the initial and final states.

The coefficients $C^{kl}$ represent the channel-dependent coupling strengths
determined by chiral symmetry constraints, with their explicit values given in
Ref.~\cite{Oset:1997it}. In that work, the potential was derived within a chiral
unitary approach and applied through a coupled-channel Lippmann–Schwinger
equation, using $f = 1.15 f_\pi$ with $f_\pi = 93$ MeV.  In the present study,
 a different framework is adopted based on the qBSE. The same chiral potential has
been used in the qBSE framework to investigate the $\Lambda(1405)$ resonance and
to describe low-energy cross sections for the processes $K^- p \to K^- p$,
$\bar{K}^0 n$, $\pi^0 \Lambda$, $\pi^0 \Sigma^0$, $\pi^+ \Sigma^-$, and $\pi^-
\Sigma^+$, yielding a good agreement with experimental data~\cite{He:2015cca}.
For consistency, this value is also adopted in the current analysis.

The rescattering amplitude ${\cal T}$, as illustrated in Fig.~\ref{diagram}(c)
and (d), can be obtained by inserting the potential kernel into the
Bethe-Salpeter equation. By employing the quasipotential approximation and
performing a partial wave decomposition, the original four-dimensional integral
equation in Minkowski space can be reduced to a one-dimensional integral
equation for each spin-parity $J^P$
channel~\cite{He:2014nya,He:2015mja,He:2017lhy,He:2015yva,He:2015cea}, as
follows,
\begin{align} i{\cal T}^{J^P}_{\lambda'\lambda}({\rm p}',{\rm p}) &=
i{\cal V}^{J^P}_{\lambda'\lambda}({\rm p}',{\rm p})+\frac{1}{2}\sum_{\lambda''}\int
\frac{{\rm p}''^2 d{\rm p}''}{(2\pi)^3} \nonumber\\ &\cdot i{\cal
V}^{J^P}_{\lambda'\lambda''}({\rm p}',{\rm p}'') G_0({\rm p}'') i{\cal
M}^{J^P}_{\lambda''\lambda}({\rm p}'',{\rm p}), \label{Eq: BS_PWA} 
\end{align}
where $\lambda$, $\lambda'$, and $\lambda''$ denote the helicities of the initial, final, and intermediate baryons, respectively. The helicities of the pseudoscalar mesons are omitted.

The propagator $G_0({\rm p}'')$ is reduced from its original four-dimensional form by the quasipotential approximation. It is given by 
\begin{align} G_0 &= \frac{\delta^+(p''^2_h - m_h^2)}{p''^2_l - m_l^2} \nonumber\\ &= \frac{\delta^+(p''^0_h - E_h({\rm p}''))}{2E_h({\rm p}'')\left[(W - E_h({\rm p}''))^2 - E_l^2({\rm p}'')\right]}, 
\end{align} 
where $p''_h$ and $p''_l$ are the four-momenta of the heavier and lighter particles in the intermediate state, respectively, and $W$ denotes the total center-of-mass energy.
In this work, the spectator approximation is adopted, where the heavier particle (denoted as $h$) is placed on-shell~\cite{Gross:1999pd}, satisfying $p''^0_h = E_h({\rm p}'') = \sqrt{m_h^2 + {\rm p}''^2}$. Consequently, the energy of the lighter particle $l$ is determined by $p''^0_l = W - E_h({\rm p}'')$. Here and throughout this paper, the three-momentum magnitude in the center-of-mass frame is denoted as ${\rm p} = |{\bm p}|$.

The partial-wave potential ${\cal V}^{J^P}_{\lambda'\lambda}({\rm p}',{\rm p})$ is obtained from the interaction potential ${\cal V}_{\lambda'\lambda}({\bm p}',{\bm p})$ through the partial wave projection as 
\begin{align} 
{\cal V}_{\lambda'\lambda}^{J^P}({\rm p}',{\rm p}) &= 2\pi\int d\cos\theta \Big[ d^{J}_{\lambda\lambda'}(\theta) {\cal V}_{\lambda'\lambda}({\bm p}',{\bm p}) \nonumber\\ & + \eta d^{J}_{-\lambda\lambda'}(\theta) {\cal V}_{\lambda'-\lambda}({\bm p}',{\bm p}) \Big], 
\end{align} 
where $d^J_{\lambda\lambda'}(\theta)$ are the Wigner $d$ functions describing the rotation between initial and final helicity states, and $\theta$ is the scattering angle between the initial and final three-momenta.
The factor $\eta$ is given by $\eta = PP_1P_2(-1)^{J - J_1 - J_2}$ where $P$ is the intrinsic parity of the system, $P_1$ and $P_2$ are the parities of the two constituent particles, and $J$, $J_1$, $J_2$ are the total and individual spins, respectively.
The initial and final three-momenta are chosen as ${\bm p} = (0,0,{\rm p})$ and ${\bm p}'= ({\rm p}'\sin\theta, 0, {\rm p}'\cos\theta)$ for convenience in performing the partial wave expansion.
In addition, an exponential form factor is introduced into the propagator to regularize the high-momentum behavior $ G_0({\rm p}'') \rightarrow G_0({\rm p}'') e^{-2(p''^2_l - m_l^2)^2/\Lambda_r^4}$ where $\Lambda_r$ is the cutoff parameter controlling the regularization strength~\cite{He:2015mja}.

The total decay amplitude corresponding to Fig.~\ref{diagram}(c) is obtained by incorporating the rescattering amplitude ${\cal T}$ into the decay process as follows:
\begin{align}
i{\cal M}_{\lambda\lambda{\Lambda_c}}(k_2,k_3) &= \sum_{\lambda'} \int \frac{d^3k'_3}{(2\pi)^3} i{\cal T}_{\lambda\lambda'}(k_2,k_3; k'_2,k'_3) \nonumber \\
&\cdot G_0(k'3) i{\cal A}_{\lambda'\lambda{\Lambda_c}}(k'_2,k'_3),
\end{align}
where $\lambda_{\Lambda_c}$, $\lambda$, and $\lambda'$ denote the helicities of the initial $\Lambda_c$, the final-state $\Lambda$, and the intermediate nucleon, respectively. The helicities of the pseudoscalar mesons are neglected, as they are not relevant for the current analysis. The momenta $k^{(')}_2$ and $k^{(')}_3$ correspond to the final (or intermediate) $\pi^+$ and $\Lambda$, respectively.
For the diagram in Fig.~\ref{diagram}(d),  $\pi^+$ should be replaced by $\pi^-$, labeled with momentum $k_4$. Other momenta not directly involved in the rescattering process are omitted for brevity.

However, since the rescattering amplitude is obtained in a partial-wave-decomposed form, a partial wave decomposition of the decay amplitude is also performed. This leads to the expression:
\begin{align}
i{\cal M}_{\lambda\lambda_{\Lambda_c}}(k_2,k_3) &= \sum_J N_J \sum_{\lambda'} \int \frac{{\rm k}'^{2}_3 d{\rm k}'_3}{(2\pi)^3} i{\cal T}^J_{\lambda\lambda'}({\rm k}'_3, M_{23})\nonumber \\
& \cdot  G_0({\rm k}'_3) i{\cal A}^{J\lambda}_{\lambda'\lambda_{\Lambda_c}}({\rm k}'_3, M_{23}),
\end{align}
where $N_J = \sqrt{(2J+1)/(4\pi)}$, and $M_{23} = \sqrt{(k_2 + k_3)^2}$ is the invariant mass of the $\pi^+ \Lambda$ system.

After including the parity quantum number, the decay amplitude can be rewritten as:
\begin{align}
i{\cal M}_{\lambda\lambda_{\Lambda_c}}(k_2,k_3) &= \sum_{J^P} \frac{N_J^2}{4} \int \frac{{\rm k}'^3_3 d{\rm k}'_3}{(2\pi)^3} \sum_{\lambda'} \nonumber \\
&\cdot i{\cal T}^{J^P}_{\lambda\lambda'}({\rm k}'_3, M_{23})G_0({\rm k}'_3)i{\cal A}^{J^P\lambda}_{\lambda'\lambda{\Lambda_c}}({\rm k}'_3, M_{23}).
\end{align}

The Lagrangian describing the initial weak decay process $\Lambda_c \to \pi \pi \bar{K} N$ is given by
\begin{align}
{\cal L}_{\Lambda_cN\pi\pi\bar{K}} = i \bar{N} \pi \pi \bar{K} (g^{PV} - g^{PC} \gamma_5) \Lambda_c,
\end{align}
where $g^{PV}$ and $g^{PC}$ denote the parity-violating and parity-conserving coupling constants, respectively. They are also taken to be equal to a common value $g^P$, which is further denoted as $g_{\Sigma(1435)^\pm}$ for consistency with the Breit-Wigner scenario.

To match the partial-wave representation used in the rescattering amplitude, the helicity amplitude for the initial $\Lambda_c$ decay is also decomposed into partial waves. The corresponding expression is
\begin{align}
{\cal A}^{J^P\lambda}_{\lambda'\lambda_{\Lambda_c}} =
\int d\Omega_j \left[ D^J_{\lambda\lambda'}(\Omega), {\cal A}_{\lambda'\lambda_{\Lambda_c}} + \eta D^J_{\lambda,-\lambda'}(\Omega) {\cal A}_{-\lambda'\lambda_{\Lambda_c}} \right],
\end{align}
where $\lambda_{\Lambda_c}$, $\lambda'$, and $\lambda$ are the helicities of the initial $\Lambda_c$, the intermediate nucleon, and the $\pi\Lambda$ system, respectively. The Wigner $D$-functions $D^J_{\lambda\lambda'}(\Omega)$ project the helicity amplitude ${\cal A}_{\lambda'\lambda_{\Lambda_c}}$ into states of definite total angular momentum $J$ and parity $P$, and $\eta$ is the intrinsic parity factor as defined above.

\section{Numerical Results}\label{Sec: results}

\subsection{Fitting procedure}

This section presents a detailed fitting analysis of the Belle data to
investigate the possible role of the $\Sigma(1/2^-)$ resonances in the decay
process $\Lambda_c \to \Lambda \pi^- \pi^+ \pi^+$. The fitting results are
summarized in Table~\ref{tab:fit}. The $\Lambda \pi^+$ invariant mass spectrum
exhibits a prominent peak around 1380 MeV, which may receive contributions from
both the $\Sigma(1385)$ resonance with spin-parity $J^P = 3/2^+$ and the
hypothesized $\Sigma(1380)$ resonance with $J^P = 1/2^-$. The $\Sigma(1385)$ is
a well-established state with a three-star rating from the PDG and is therefore
included by default in all fitting scenarios to ensure a consistent description
of the dominant feature in the spectrum.

\renewcommand\tabcolsep{0.38cm}
\renewcommand{\arraystretch}{1.9}
\begin{table*}[h!]

  \caption{ Fitting parameters and $\chi^2$ values for different scenarios
  labeled as $(ij)$, where $i$ and $j$ indicate the inclusion ($1$) or exclusion
  ($0$) of the $\Sigma(1380)$ and $\Sigma(1435)$ resonances, respectively. The
  subscript ``BW" refers to cases where the $\Sigma(1435)$ is treated as a
  Breit-Wigner resonance, while ``RS" denotes a rescattering model. The
  superscript ``p" indicates fits performed using a subset of the data.  The
  strength parameters $g$ for the resonances, rescattering contribution, and
  background (bk) are scaled to match the overall event yield; therefore, their
  absolute values have no physical units and only their relative magnitudes are
  meaningful. The phase parameters $\phi$ are dimensionless. The background
  shape parameters $c_\pm$ are given in units of GeV$^{-1}$, and the masses $m$
  and widths $\Gamma$ are in units of MeV.  The $\chi^2_-$ and $\chi^2_+$ values
  correspond to the fits for the $\Lambda\pi^-$ and $\Lambda\pi^+$ invariant
  mass spectra, respectively. The total fit quality is indicated by $\chi^2_{\rm
  tot}$, along with the total number of data points and the number of degrees of
  freedom (ndf).  \label{tab:fit}}
  
    \begin{tabular}{c|rrrrrrrr}\toprule[2pt]
                   Fits                &$(00)$    &$\rm (01_{BW})^p$ & $\rm (01_{BW})$ & $\rm (01_{RS})$ &$\rm (11_{BW})$ & $\rm (11_{RS})$ & $(10)$ & Reference\\\hline
$g_{\Sigma(1380)^-}$               &$--$       &$--$           &$--$          &$--$          &$0.0000$     &$0.0149$      &$0.0000$      \\
$\phi_{\Sigma(1380)^-}$            &$--$       &$--$           &$--$          &$--$          &$0.0231$     &$0.5611$      &$0.0297$       \\
$g_{\Sigma(1380)^+}$               &$--$       &$--$           &$--$          &$--$          &$0.0151$     &$0.0107$      &$0.0161$       \\
$\phi_{\Sigma(1380)^+}$            &$--$       &$--$           &$--$          &$--$          &$0.2706$     &$0.4282$      &$0.2656$       \\\hline
$g_{\Sigma(1385)^-}$               &$0.3016$   &$0.2666$       &$0.2984$      &$0.3259$      &$0.2976$     &$0.3356$      &$0.2993$   \\
$\phi_{\Sigma(1385)^-}$            &$0.7290$   &$0.7211$       &$0.7301$      &$0.7765$      &$0.7276$     &$0.7229$      &$0.7187$    \\
$m_{\Sigma(1385)^-}$               &$1384.8$   &$1386.6$       &$1386.1$      &$1384.3$      &$1384.7$     &$1385.3$      &$1384.7$   &$1387.2\pm0.5$~\cite{ParticleDataGroup:2024cfk}  \\
$\Gamma_{\Sigma(1385)^-}$          &$34.6$     &$30.8$         &$34.5$        &$36.5$        &$33.0$       &$37.1$      &$33.8$   &$39.4\pm2.1$~\cite{ParticleDataGroup:2024cfk}    \\
$g_{\Sigma(1385)^+}$               &$0.4824$   &$0.4662$       &$0.4640$      &$0.4146$      &$0.3224$     &$0.3645$      &$0.3313$      \\
$\phi_{\Sigma(1385)^+}$            &$0.2677$   &$0.2727$       &$0.2738$      &$0.4052$      &$0.2763$     &$0.2741$      &$0.2779$      \\
$m_{\Sigma(1385)^+}$               &$1378.6$   &$1378.4$       &$1378.8$      &$1380.4$      &$1380.5$     &$1381.1$      &$1380.7$   &$1382.83\pm0.34$~\cite{ParticleDataGroup:2024cfk}   \\
$\Gamma_{\Sigma(1385)^+}$          &$41.0$     &$39.3$         &$40.0$        &$36.8$        &$30.1$       &$34.0$      &$31.4$   &$36.2\pm0.7$~\cite{ParticleDataGroup:2024cfk}     \\\hline
$g_{\Sigma(1435)^-}$               &$--$       &$0.00043$      &$0.00037$     &$8.5904$      &$0.00037$    &$9.6326$      &$--$        \\
$\phi_{\Sigma(1435)^-}$            &$--$       &$0.1883$       &$0.1253$      &$0.7513$      &$0.1539$     &$0.8291$      &$--$        \\
$m_{\Sigma(1435)^-}$               &$--$       &$1434.7$       &$1433.5$      &$--$          &$1433.5$     &$--$          &$--$    & $1438.5\pm0.9$~\cite{Belle:2022ywa}         \\
$\Gamma_{\Sigma(1435)^-}$          &$--$       &$11.9$         &$11.5$        &$--$          &$11.5$       &$--$          &$--$    & $33.0\pm7.5$~\cite{Belle:2022ywa}          \\
$g_{\Sigma(1435)^+}$               &$--$       &$0.00093$      &$0.00097$     &$7.515$      &$0.00066$    &$7.2618$      &$--$       \\
$\phi_{\Sigma(1435)^+}$            &$--$       &$0.2755$       &$0.3354$      &$0.7577$      &$0.3510$     &$0.8252$      &$--$        \\
$m_{\Sigma(1435)^+}$               &$--$       &$1435.1$       &$1438.0$      &$--$          &$1438.0$     &$--$          &$--$     &$1434.3\pm0.6$~\cite{Belle:2022ywa}         \\
$\Gamma_{\Sigma(1435)^+}$          &$--$       &$14.2$         &$18.5$        &$--$          &$12.6$       &$--$          &$--$     &$11.5\pm2.8$~\cite{Belle:2022ywa}        \\\hline
$g_{\rm bk}$                           &$1.3012$   &$1.3010$       &$1.3009$      &$1.1860$      &$1.2894$     &$1.1896$      &$1.2894$         \\
$c_-$                              &$0.0217$   &$0.0206$       &$0.0224$      &$0.0045$      &$-0.0135$    &$0.0243$      &$-0.0104$        \\
$c_+$                              &$-0.0870$  &$-0.0867$      &$-0.0880$     &$-0.0872$     &$-0.1081$    &$-0.0900$     &$-0.1080$    \\\bottomrule[1pt]
${\chi^2_-}/{N_{\Lambda\pi^-}}$    &$276/100$  &$97/75$        &$156/100$     &$216/100$     & $156/100$   &$137/100$     &$215/100$    &\\
${\chi^2_+}/{N_{\Lambda\pi^+}}$    &$556/100$  &$106/75$       &$262/100$     &$235/100$     & $153/100$   &$154/100$     &$294/100$    \\
${\chi^2_{\rm tot}}/{N_{\rm tot}}$ &$832/200$  &$203/150$      &$418/200$     &$451/200$     & $309/200$   &$291/200$     &$509/200$    \\
${\chi^2_{\rm tot}}/{\rm ndf}$     &$4.40$     &$1.55$         &$2.31$        &$2.43$        & $1.75$      &$1.61$        &$2.81$    \\
\bottomrule[2pt]
\end{tabular}
\end{table*}

In contrast, the $\Sigma(1380)$ is less well established and is typically considered a broad resonance in the literature. Due to its large width, its contribution tends to be smeared out in the spectrum, resulting in relatively low sensitivity to its exact mass and width during the fitting procedure. To avoid introducing unnecessary free parameters, its mass and width are fixed to 1380  and 120~MeV, respectively, following the values adopted in previous studies~\cite{Wu:2009tu,Wu:2009nw,Liu:2017hdx,Lyu:2024qgc}.

Given the dominant contribution of the $\Sigma(1385)$ resonance to the spectrum, the analysis begins with a baseline fit, labeled (00) in Table~\ref{tab:fit}, which includes only the $\Sigma(1385)$ and a nonresonant background component. To further investigate the origin of the structure observed around 1435 MeV, the $\Sigma(1435)$ resonance is introduced into the fitting framework. Since the Belle Collaboration applied a selection cut to enhance the signal in their analysis, two corresponding fits are performed: one using the partial data points between 1.38 and 1.53 GeV, denoted as $\rm  (01_{BW})^p$, and another using the full data set, denoted as $(01_{\rm BW})$. In both cases, the $\Sigma(1435)$ is treated as a Breit-Wigner resonance.

To explore the possibility that the $\Sigma(1435)$ is not a conventional Breit-Wigner resonance but rather a dynamically generated state arising from coupled-channel interactions, an additional fit is performed in which the $\Sigma(1435)$ contribution is modeled via a rescattering mechanism. This fit is referred to as $(01_{\rm RS})$.

The impact of the $\Sigma(1380)$ resonance is further examined by including both the $\Sigma(1385)$ and $\Sigma(1380)$ in the model. When the $\Sigma(1435)$ is incorporated as a Breit-Wigner resonance, the corresponding fit is labeled as $(11_{\rm BW})$; when it is instead treated through rescattering, the fit is denoted as $(11_{\rm RS})$. To isolate the role of the $\Sigma(1380)$, another fit is carried out that includes the $\Sigma(1385)$ and $\Sigma(1380)$, but excludes the $\Sigma(1435)$ entirely; this scenario is labeled (10).

The fitted parameters are summarized in Table~\ref{tab:fit}. For the $\Sigma(1380)$ resonance, the mass and width are fixed during the fit, and only the coupling strengths $g_{\Sigma(1380)^\pm}$ and the relative phases $\phi_{\Sigma(1380)^\pm}$ are treated as free parameters. In cases where the $\Sigma(1385)$ and $\Sigma(1435)$ are modeled as Breit-Wigner resonances, their masses $m$ and widths $\Gamma$ are also treated as fit parameters, in addition to their respective strengths and phases.
To account for nonresonant background contributions, three additional parameters are introduced: the overall strength $g_{\rm bk}$ and the shape parameters $c_-$ and $c_+$. A common normalization factor is applied to all resonance and background strengths $g$ to ensure consistency with the total event yield observed in the experimental data. Consequently, only the relative magnitudes of these parameters carry physical significance, while their absolute values are not meaningful.
The quality of each fit is evaluated using the corresponding $\chi^2$ values. Specifically, $\chi^2_-$ and $\chi^2_+$ are reported for the $\Lambda\pi^-$ and $\Lambda\pi^+$ invariant mass distributions, respectively, each accompanied by the number of data points used. The total $\chi^2_{\rm tot}$, along with the total number of data points and the number of degrees of freedom, is also provided to quantify the overall fit quality.

\subsection{Role of $\Sigma(1435)$}

\begin{figure}[h!]
  \includegraphics[bb=5 0 1000 490,clip,scale=0.69]{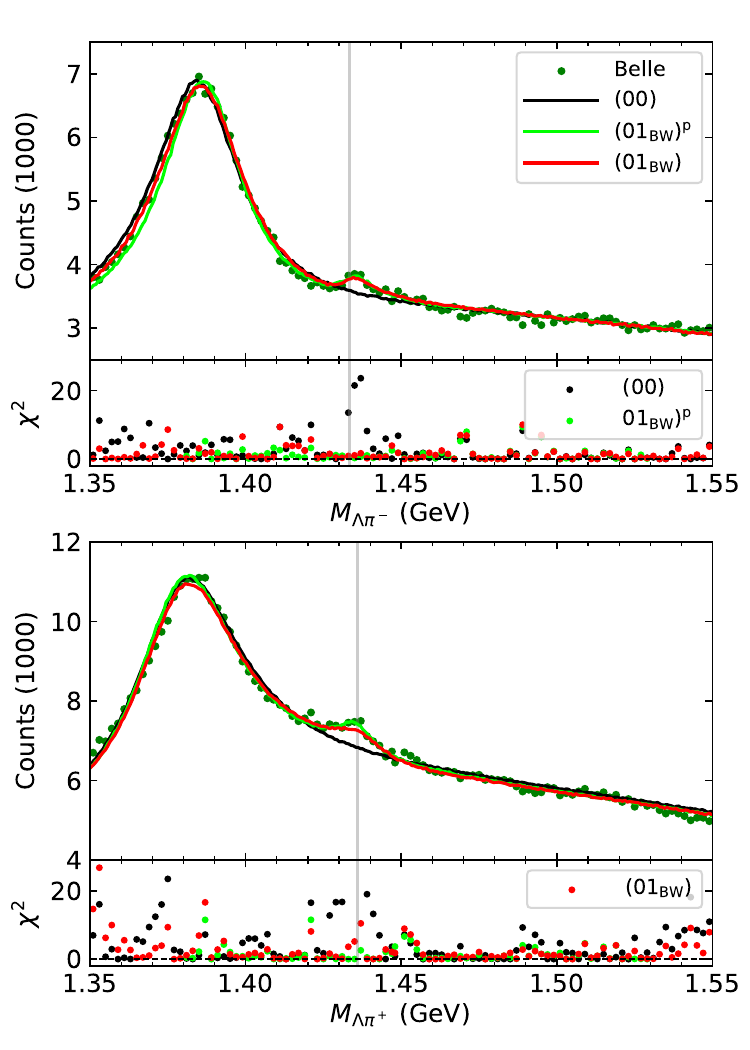}
  \caption{Invariant mass spectra of $\Lambda\pi^-$ (upper panel) and $\Lambda\pi^+$ (lower panel). The curves correspond to the fit results for three scenarios: $(00)$ including only the $\Sigma(1385)$ and background (black); $(01_{\rm BW})^p$ including the $\Sigma(1435)$ and using partial data (green); and $(01_{\rm BW})$ including the $\Sigma(1435)$ with the full data set (red). The data are taken from the Belle Collaboration~\cite{Belle:2022ywa}. The vertical grey lines indicate the $K^-n$ and $\bar{K}^0p$ thresholds. The scatter points in the lower part of each panel represent the individual $\chi^2$ contributions of the corresponding data points. Legends apply to both panels.}
  \label{2}
\end{figure}

Figure~\ref{2} shows the fits to the $\Lambda\pi^\pm$ invariant mass spectra with and without the inclusion of the $\Sigma(1435)$ resonance. Three fitting schemes are considered: (00), $\rm (01_{BW})^p$, and $(01_{\rm BW})$. In fit (00), only the $\Sigma(1385)$ with spin-parity $J^P = 3/2^+$ and a non-resonant background are included. This minimal setup is sufficient to reproduce the overall structure observed in the Belle data for both the $\Lambda\pi^-$ and $\Lambda\pi^+$ channels.

The corresponding $\chi^2$ values for each data point are also provided under
fit (00). In the $\Lambda\pi^-$ spectrum, the majority of the data points
exhibit small $\chi^2$ values, indicating a good fit in general. However,
significant discrepancies are observed in the vicinity of 1435~MeV, where the
$\chi^2$ values are notably larger. A few data points near the lower mass region
around 1350~MeV also show elevated $\chi^2$. Similarly, in the $\Lambda\pi^+$
spectrum, large $\chi^2$ values are concentrated around 1435~MeV and extend down
to the 1350–1380~MeV range. Additional discrepancies are observed in the
high-mass region near 1.55~GeV. The theoretical curves for invairnat mass
spectrum show clear deviations from the experimental data, particularly around
1345~MeV.

The clear deviation from experimental data around 1435~MeV suggests the presence
of a resonance with a mass near 1435~MeV. In Fig.~\ref{2}, the results of the
fit $(01)_{\rm BW}$, which includes the $\Sigma(1435)$ modeled as a Breit-Wigner
resonance, are presented. A substantial reduction in the $\chi^2$ value is
observed, indicating a significant improvement in the fit quality—particularly
in the region near 1435~MeV—where the model now better reproduces the
experimental event distribution.  As summarized in Table~\ref{tab:fit}, the
total $\chi^2$ values for the $\Lambda\pi^-$ and $\Lambda\pi^+$ spectra in the
baseline fit (00) are 276 and 556, respectively, leading to a combined
$\chi^2_{\rm tot}/{\rm ndf}$ of 4.40. When the $\Sigma(1435)$ is included as a
Breit-Wigner resonance, the total $\chi^2_{\rm tot}/{\rm ndf}$ decreases
significantly to 2.31, with the individual contributions reduced to 156 for the
$\Lambda\pi^-$ spectrum and 263 for the $\Lambda\pi^+$ spectrum.  The relatively
larger $\chi^2$ value for the $\Lambda\pi^-$ channel mainly arises from
discrepancies in the lower energy region, where the fit exhibits poorer
agreement with the experimental data. 

Results based on the partial dataset in the range 1.38-1.53~GeV are also
presented, following the strategy adopted in the Belle analysis, which
specifically targets the $\Sigma(1435)$ region. In this case, the $\chi^2$
values for the $\Lambda\pi^-$ and $\Lambda\pi^+$ spectra are 97 and 106,
respectively, comparable to the Belle Collaboration's reported values of 74 and
93~\cite{Belle:2022ywa}. The present analysis employs 19 fit parameters, a
number similar to the 18 used in the Belle study. Furthermore, both the
$\Lambda\pi^-$ and $\Lambda\pi^+$ spectra are fitted simultaneously, with
interference effects among different contributions explicitly taken into
account. As such, the current fitting procedure is comparable to the Belle
analysis in terms of both complexity and reliability. Notably, even when the
full dataset is used, the average $\chi^2$ remains close to that obtained from
the partial dataset, further demonstrating the robustness of the fit.

The masses and widths of both the $\Sigma(1385)$ and $\Sigma(1435)$ are fitted in the $\rm (01_{BW})^p$ and $\rm (01_{BW})$ scenarios. For the $\Sigma(1385)$, the fitted mass and width for the negatively charged state are 1386.6 MeV and 30.8 MeV in the $\rm (01_{BW})^p$ fit, and 1386.1 MeV and 34.5 MeV in the $\rm (01_{BW})$ fit. For the positively charged state, the corresponding values are 1378.4 MeV and 39.3 MeV in $\rm (01_{BW})^p$, and 1378.8 MeV and 40.0 MeV in $\rm (01_{BW})$. These values are consistent with those reported by the PDG~\cite{ParticleDataGroup:2024cfk}.
For the $\Sigma(1435)$, the fitted mass and width are 1434.7  and 11.9 MeV for the negatively charged state, and 1435.1  and 14.2 MeV for the positively charged state. Compared with the measurements reported by the Belle Collaboration~\cite{Belle:2022ywa}, the widths of the two charge states in our fit show better consistency with each other.

\subsection{Breit-Wigner vs rescattering  for $\Sigma(1435)$ structure}
 
The $\Sigma(1385)$ provides the dominant contribution in the energy region covered by the Belle experiment. Once the $\Sigma(1435)$ structure is included as a Breit-Wigner resonance, the experimental data are well reproduced. Two scenarios are considered for modeling the $\Sigma(1435)$: one treats it as a Breit-Wigner resonance, as implemented in the fit labeled $(01_{\rm BW})$, and the other describes it via a rescattering mechanism, corresponding to the fit $(01_{\rm RS})$. The results for the rescattering scenario are summarized in Table~\ref{tab:fit}.
Compared with the baseline fit $(00)$, which includes only the $\Sigma(1385)$ and background contributions, the rescattering scenario yields a substantial reduction in the total $\chi^2$, decreasing from 832 to 451 for the full dataset. Although this is slightly higher than the value obtained in the Breit-Wigner case, it still represents a significant improvement in fit quality. Notably, the fitted parameters for the $\Sigma(1385)$ remain nearly unchanged across the different scenarios. No explicit resonance parameters are associated with the $\Sigma(1435)$ in the rescattering fit, since its contribution arises dynamically rather than from a predefined Breit-Wigner form.

\begin{figure}[h!]
    \includegraphics[bb=5 18 1000 640,clip,scale=0.69]{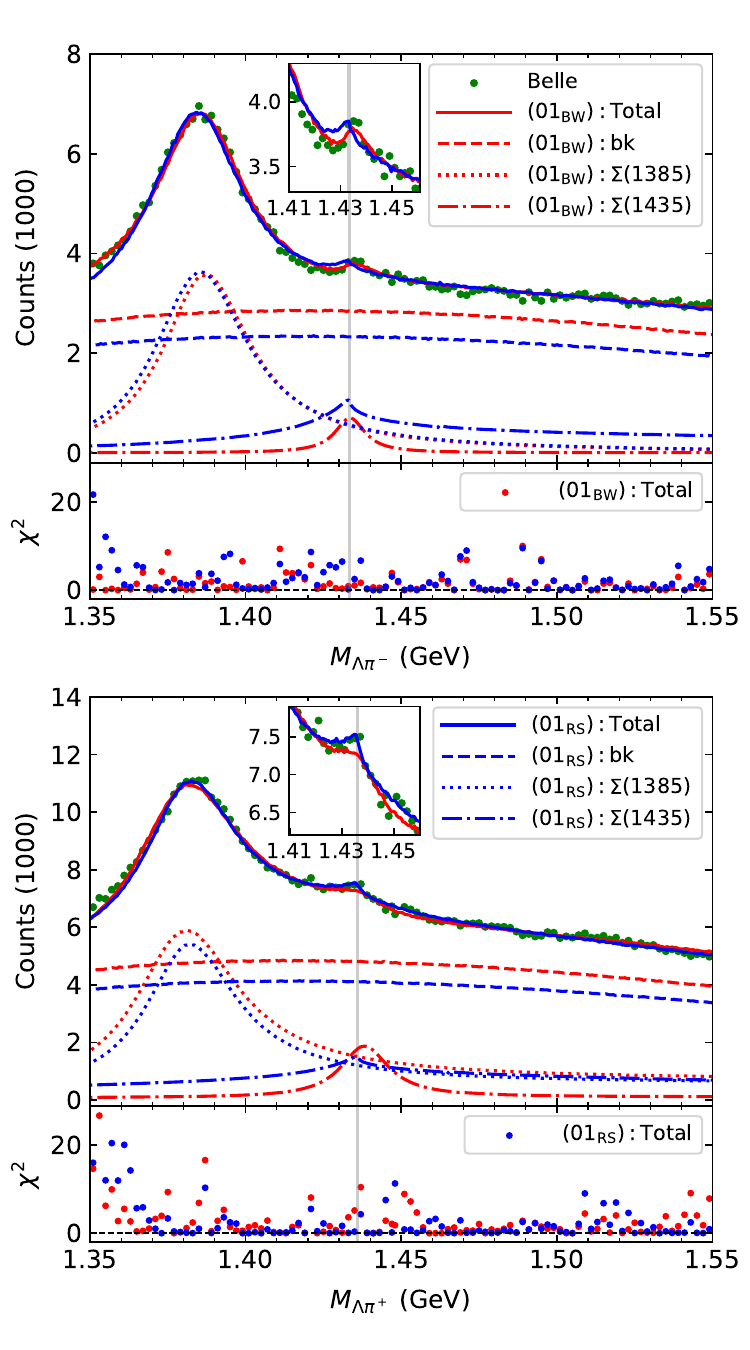}
    \caption{Invariant mass spectra of $\Lambda\pi^-$ (upper panel) and $\Lambda\pi^+$ (lower panel) with the inclusion of the $\Sigma(1435)$ resonance. The curves correspond to two fitting scenarios: $(01_{\rm BW})$, where the $\Sigma(1435)$ is modeled as a Breit-Wigner resonance (red), and $(01_{\rm RS})$, where the $\Sigma(1435)$ arises from a rescattering mechanism (blue). In each case, the solid, dashed, dotted, and dash-dotted lines represent the total, background, $\Sigma(1385)$, and $\Sigma(1435)$ contributions, respectively. For visibility, the $\Sigma(1435)$ contribution in the $\Lambda\pi^-$ invariant mass spectrum (upper panel) is scaled by a factor of 100. The insets in both the upper and lower panels show a magnified view of the region around 1435MeV. Experimental data are taken from the Belle Collaboration~\cite{Belle:2022ywa}. The vertical grey lines indicate the $K^-n$ and $\bar{K}^0p$ thresholds. The scatter points at the bottom of each panel show the $\chi^2$ values for individual data points. Legends apply to both panels.}
    \label{3}
\end{figure}

As shown in Fig.~\ref{3}, the fitted invariant mass spectra from both scenarios are nearly indistinguishable and provide a good description of the experimental data, except in the low-energy region of the $\Lambda\pi^+$ spectrum, where some discrepancies persist. The individual contributions from the $\Sigma(1385)$, $\Sigma(1435)$, and background are also displayed in Fig.~\ref{3}. Although the total fit results are similar for both mechanisms, the decomposition into individual components reveals notable differences. In both cases, the background contributes significantly across the entire energy range and becomes dominant at higher invariant masses, with a larger contribution observed in fit $\rm (01_{BW})$. At lower energies, the $\Sigma(1385)$ dominates the spectrum, and its contribution is comparable between the two scenarios.

The $\Sigma(1435)$ structure appears around 1435 MeV as expected, either as a resonance or through rescattering. Its contribution is substantially more pronounced in the rescattering scenario than in the Breit-Wigner case. In fact, in the Breit-Wigner fit, the direct contribution from the $\Sigma(1435)$ is relatively small and has been scaled by a factor of 100 for visibility. Unlike the Breit-Wigner component, which is sharply localized near 1435 MeV, the rescattering contribution spans a broader energy range and effectively compensates for the reduced background in fit $(01_{\rm RS})$. Interestingly, although the direct Breit-Wigner amplitude is minimal, it still induces a noticeable enhancement in the total event yield near 1435 MeV, primarily due to interference effects with other components.

Furthermore, the rescattering mechanism generates a clear cusp precisely at the $K^-n$ and $\bar{K}^0p$ thresholds, whereas the peak of the Breit-Wigner resonance lies slightly above these thresholds. As shown in the inset of Fig.~\ref{3}, the $\Lambda\pi^+$ invariant mass spectrum measured by Belle exhibits a distinct cusp-like structure, which strongly favors the rescattering interpretation—especially in light of the sharp drop observed at the threshold. In contrast, the $\Lambda\pi^-$ spectrum shows no evident cusp behavior.

It is also worth noting that, unlike the Belle analysis, where separate cusp parameters were introduced independently for the two charge states, the present study incorporates an explicit interaction mechanism that predicts a common cusp shape for both the $\Lambda\pi^+$ and $\Lambda\pi^-$ channels. This constraint results in a poorer fit to the $\Lambda\pi^-$ spectrum compared to the Belle fit and the Breit-Wigner scenario $(01_{\rm BW})$.

\subsection{Role of $\Sigma(1380)$}

Although the $\Lambda\pi^\pm$ invariant mass spectra are generally well
reproduced with the inclusion of the $\Sigma(1435)$, noticeable discrepancies
remain in the low-energy region, particularly in the $\Lambda\pi^+$ spectrum. In
this region, the existence of a $\Sigma(1380)$ state has been proposed in the
literature. Previous studies have demonstrated that the combined inclusion of
the $\Sigma(1380)$ and $\Sigma(1385)$ is essential to account for the
near-threshold enhancement observed in the $p\bar{K}^0$ invariant mass spectrum
of the $\Lambda_c^+ \to p \bar{K}^0 \eta$ decay. Motivated by these findings,
the present analysis explores the potential impact of the $\Sigma(1380)$ by
incorporating it into fits $(11_{\rm BW})$ and $(11_{\rm RS})$. Since the
$\Sigma(1380)$ does not generate a pronounced structure in the invariant mass
distributions, it is modeled as a Breit-Wigner resonance for simplicity in both
cases.

\begin{figure}[h!]
  \includegraphics[bb=5 18 1000 640,clip,scale=0.69]{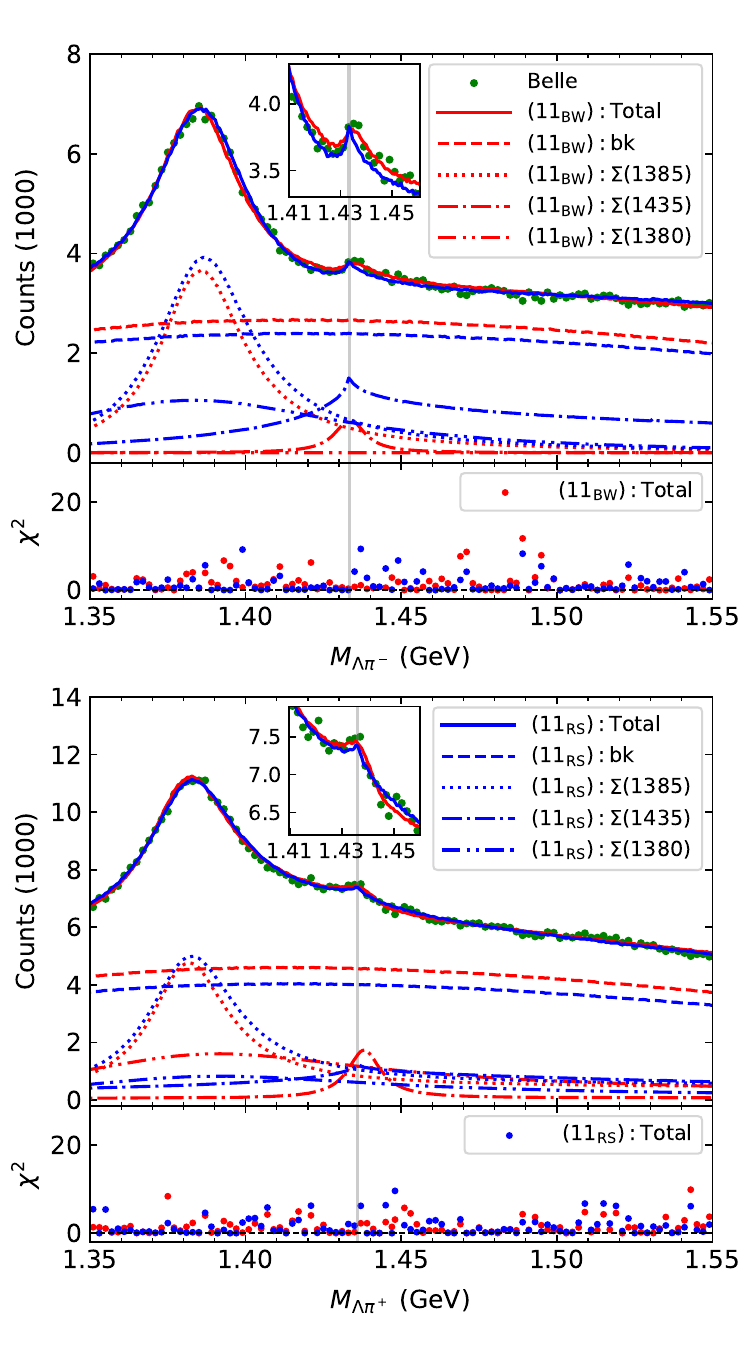}
  \caption{Invariant mass spectra of $\Lambda\pi^-$ (upper panel) and $\Lambda\pi^+$ (lower panel) with the inclusion of both the $\Sigma(1380)$ and $\Sigma(1435)$ resonances. The curves correspond to two fitting scenarios: $(11_{\rm BW})$, where the $\Sigma(1435)$ is modeled as a Breit-Wigner resonance (red), and $(01_{\rm RS})$, where the $\Sigma(1435)$ is dynamically generated via rescattering (blue). In each case, the solid, dashed, dotted, dash-dotted, and dash-dot-dotted curves represent the total, background, $\Sigma(1385)$, $\Sigma(1435)$, and $\Sigma(1380)$ contributions, respectively. For visibility, the $\Sigma(1435)$ contribution in the $\Lambda\pi^-$ invariant mass spectrum (upper panel) is scaled by a factor of 100, and the $\Sigma(1380)$ contributions in both panels are scaled by a factor of 10. Insets in both panels provide a magnified view of the region around 1435MeV. The experimental data are taken from the Belle Collaboration~\cite{Belle:2022ywa}. Vertical grey lines indicate the $K^-n$ and $\bar{K}^0p$ thresholds. The scatter points at the bottom of each panel represent the $\chi^2$ contributions from individual data points. Legends apply to both panels.}
  \label{4}
\end{figure}

With the inclusion of the $\Sigma(1380)$, the fitting in the low-energy region is significantly improved, as the $\chi^2$ for the data points shown in Fig.~\ref{3} decreases substantially. It can be observed that the $\chi^2$ values in the low-energy region for both $\Lambda\pi^-$ and $\Lambda\pi^+$ invariant mass spectra are almost zero, as shown in Fig.~\ref{4}. This improvement is observed in both fittings with $\Sigma(1435)$ as a Breit-Wigner resonance and as rescattering. For the former, the total $\chi^2$ decreases from 418 to 309, while for the latter, it drops from 451 to 291.

The explicit contributions are also presented in Fig.~\ref{4}. The background contributions are similar to those without the $\Sigma(1380)$, though the fitting $\rm (11_{BW})$ has a slightly larger background contribution compared to $\rm (11_{RS})$. Combined with the background, the contribution from $\Sigma(1385)$ remains dominant in the lower-energy region. However, the contributions from $\Sigma(1380)$ differ between the Breit-Wigner and rescattering fittings. In the $\rm (11_{BW})$ fitting, the $\Sigma(1380)$ plays a negligible role in the $\Lambda\pi^-$ spectrum but has a significant impact in the lower-energy region of the $\Lambda\pi^+$ spectrum. In contrast, in the $\rm (11_{RS})$ fitting, the $\Sigma(1380)$ plays an important role in both spectra, with a larger contribution in the $\Lambda\pi^-$ spectrum. The contributions from $\Sigma(1435)$ as a Breit-Wigner resonance and as rescattering are similar to those without the $\Sigma(1380)$, but the total fit for the structure around 1435~MeV is clearly improved.

In the $\Lambda\pi^-$ spectrum, both scenarios for $\Sigma(1435)$ can produce the dip before the peak, which is not reproduced in the rescattering case without $\Sigma(1380)$. This improvement is attributed to the interference between $\Sigma(1380)$ and rescattering contributions. However, the threshold for $K^-n$ is slightly lower than the experimental peak, making the fit with rescattering slightly worse than the Breit-Wigner resonance. For the $\Lambda\pi^+$ spectrum, both models provide a good fit, with the Breit-Wigner resonance fitting further improved by the inclusion of the $\Sigma(1380)$.

\begin{figure}[h!]
  \includegraphics[bb=0 20 1000 490,clip,scale=0.48]{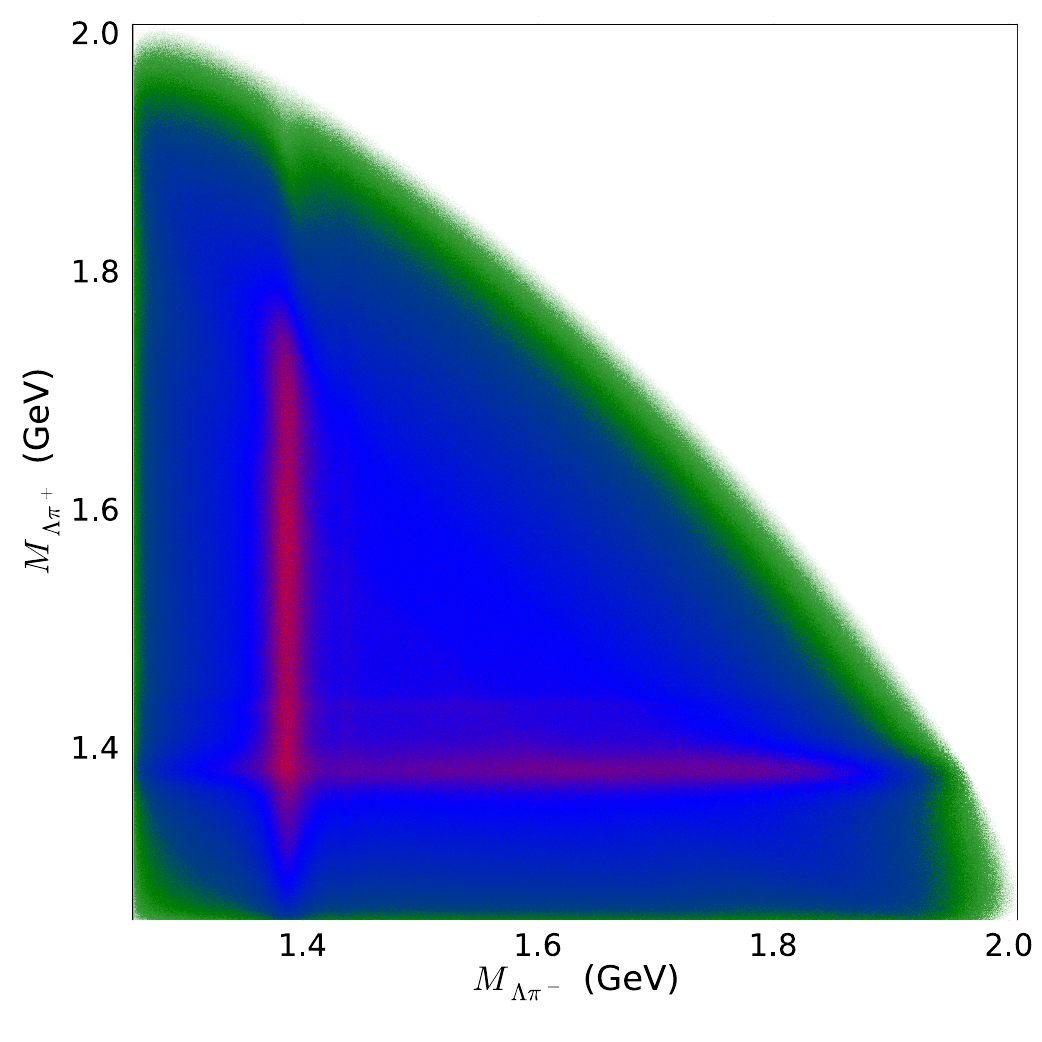}
  \caption{Dalitz-like plot for $\Lambda_c \to \Lambda \pi^+ \pi^+ \pi^-$.}
  \label{DP}
\end{figure}

Figure~\ref{DP} shows a Dalitz-like plot corresponding to the fit $(11_{\rm BW})$, provided here for reference. A similar distribution is obtained for the fit $(11_{\rm RS})$.  It can be observed that two strips are clearly parallel to the vertical and horizontal axes for the $\Lambda\pi^+$ and $\Lambda\pi^-$ invariant mass spectra, corresponding to the $\Sigma(1385)$ and the interference with $\Sigma(1380)$. Additionally, two dim strips near the main strips represent the contributions from the $\Sigma(1435)$.

\section{Summary}\label{Summary}

In this work, the Belle data for the decay process $\Lambda_c \to \Lambda \pi^+
\pi^+ \pi^-$ are analyzed using effective Lagrangians, meson-baryon rescattering
mechanisms, and the qBSE approach, in
combination with event simulations. Simulated events are used to generate the
$\Lambda \pi^+$ and $\Lambda \pi^-$ invariant mass spectra, which are
subsequently fitted to the experimental data. The model includes contributions
from the well-established $\Sigma(1385)$ resonance, two spin-parity $1/2^-$
$\Sigma$ states—particularly the hypothesized $\Sigma(1380)$—as well as
nonresonant background processes. To describe the structure observed near 1435
MeV, two scenarios are considered: one in which the $\Sigma(1435)$ is modeled as
a Breit-Wigner resonance, and another in which it emerges dynamically through
coupled-channel rescattering. The combined contributions from these components
allow for a satisfactory reproduction of the experimental spectra.

The $\Sigma(1385)$, together with background contributions, provides the dominant contribution in both the $\Lambda \pi^+$ and $\Lambda \pi^-$ spectra. However, to account for the enhancement observed near 1435 MeV, an additional mechanism is necessary. The inclusion of the $\Sigma(1435)$—either as a Breit-Wigner resonance or through rescattering—helps to describe this structure. While the Breit-Wigner scenario can reproduce the general shape of the enhancement and leads to an improved $\chi^2$ relative to the case without it, the data exhibit a cusp-like feature near the $K^- n$ and $\bar{K}^0 p$ thresholds, which is more naturally explained by the rescattering mechanism. This cusp behavior, evident in the $\Lambda \pi^+$ spectrum, supports the interpretation of the $\Sigma(1435)$ as a dynamically generated structure.

Despite these improvements, discrepancies remain in the low-energy region of the $\Lambda \pi^+$ spectrum, which neither the BW nor rescattering scenario alone can fully resolve. To address this, the $\Sigma(1380)$ resonance is included with spin-parity $1/2^-$, as suggested in the literature. Although this state does not manifest as a distinct peak in the invariant mass spectra, its inclusion as a broad Breit-Wigner resonance significantly improves the description of the low-energy region and enhances the fit quality near 1435 MeV as well. Consequently, the combined inclusion of the $\Sigma(1380)$ and $\Sigma(1435)$ leads to a markedly improved agreement with the experimental data.

To quantitatively assess the importance of the intermediate states, their statistical significance is evaluated based on the $\chi^2$ values presented in Table~\ref{tab:fit}. Specifically, comparisons are made between the full model and variants excluding either the $\Sigma(1435)$ or $\Sigma(1380)$ resonances. The resulting significance values under both scenarios are summarized in Table~\ref{tab:Sig}. In all cases, both the $\Sigma(1435)$ and $\Sigma(1380)$ exhibit high statistical significance, indicating their essential roles in accurately reproducing the experimental spectra. The significance of the $\Sigma(1435)$ is comparable between the two scenarios, while the $\Sigma(1380)$ demonstrates slightly higher significance in the rescattering scenario, suggesting that its contribution becomes more critical when the $\Sigma(1435)$ is modeled as a dynamically generated state.
\renewcommand\tabcolsep{0.32cm}
\renewcommand{\arraystretch}{1.8}
\begin{table}[h!]
  \caption{Statistical significance of the $\Sigma(1435)$ and $\Sigma(1380)$ contributions in the fits under two scenarios: treating the $\Sigma(1435)$ as a Breit-Wigner (BW) resonance or as a rescattering (RS) effect. \label{tab:Sig}}
\begin{tabular}{c|cccc}\toprule[2pt]
        $\frac{\Sigma^*}{\rm scenario}$       &$\frac{\Sigma(1435)}{\rm BW}$ &$\frac{\Sigma(1380)}{\rm BW}$  & $\frac{\Sigma(1435)}{\rm RS}$ &$\frac{\Sigma(1380)}{\rm RS}$\\\hline
Significance ($\sigma$)  &$13.0$     & $9.7$   &$14.2$     &$12.0$    \\
\bottomrule[2pt]
\end{tabular}
\end{table}

In conclusion, the combined contributions from the $\Sigma(1385)$ and $\Sigma(1435)$, treated both as a Breit-Wigner resonance and via rescattering, and the $\Sigma(1380)$ resonance provide an effective description of the $\Lambda_c \to \Lambda \pi^+ \pi^+ \pi^-$ decay process. This comprehensive model yields a more accurate reproduction of the full experimental invariant mass spectra and highlights the essential role of interference effects among the resonances in describing the observed data.

\vskip 10pt \noindent {\bf Acknowledgement} I would like to thank Professor Kiyoshi Tanida for kindly explaining the uncertainties of the data and the statistical significance. This project is supported by the National Natural Science Foundation of China (Grant No. 12475080)

\end{document}